\documentclass[aps,prd,onecolumn,superscriptaddress,nofootinbib,showpacs]{revtex4}

\usepackage{graphicx}
\usepackage{slashed}
\usepackage{amsmath,bbm,latexsym,amssymb}
\usepackage{epsfig}
\usepackage{epstopdf}

\newcommand{\be}{\begin{equation}}
\newcommand{\ee}{\end{equation}}
\newcommand{\ba}{\begin{eqnarray}}
\newcommand{\ea}{\end{eqnarray}}
\newcommand{\bs}{\begin{subequations}}
\newcommand{\es}{\end{subequations}}

\newcommand{\ra}{\rightarrow}
\newcommand{\vp}{\varphi}

\newcommand{\grts}{\raise.3ex\hbox{$>$\kern-.75em\lower1ex\hbox{$\sim$}}}
\newcommand{\lets}{\raise.3ex\hbox{$<$\kern-.75em\lower1ex\hbox{$\sim$}}}

\begin{document}
\vspace*{1cm}

\title{Two Higgs doublet models with an $S_3$ symmetry}

\author{D.\ Cogollo}\thanks{E-mail: diego.cogollo@cftp.ist.utl.pt}
\affiliation{CFTP, Departamento de F\'{\i}sica,
Instituto Superior T\'{e}cnico, Universidade de Lisboa,
Avenida Rovisco Pais 1, 1049 Lisboa, Portugal}
\affiliation{Departamento de F\'{\i}sica, Universidade Federal de Campina Grande,
Caixa Postal 10071, 58109-970, Campina Grande, PB, Brazil}
\author{Jo\~{a}o P.\ Silva}\thanks{E-mail: jpsilva@cftp.ist.utl.pt}
\affiliation{CFTP, Departamento de F\'{\i}sica,
Instituto Superior T\'{e}cnico, Universidade de Lisboa,
Avenida Rovisco Pais 1, 1049 Lisboa, Portugal}

\date{\today}

\begin{abstract}
We study all implementations of the $S_3$ symmetry in the two Higgs doublet model
with quarks, consistent with non-zero quark masses and a
Cabibbo-Kobayashi-Maskawa (CKM) matrix which is not block diagonal.
We study the impact of the various soft-breaking terms and vacuum
expectation values,
and find an interesting relation between
the mixing angles
$\alpha$
and $\beta$.
We also show that,
in this minimal setting,
only two types of assignments are possible:
either all field sectors are in singlets or all field sectors
have a doublet.
\end{abstract}


\maketitle

\section{\label{sec:intro}Introduction}

Since the discovery of a spin 0 particle at LHC \cite{ATLASHiggs, CMSHiggs}
a crucial task in particle physics has been the determination
of the number of such particles.
In the Standard Model (SM), the existence
of a single scalar doublet is a feature
without a strong theoretical justification.
As a result,
multi scalar models have been studied for a long time,
in particular, two Higgs doublet models (2HDM)
\cite{hhg, ourreview}.

Because such models involve many new parameters,
one may curtail their number by imposing extra
symmetries.
The identification of all possible symmetries
that may be imposed on the scalar potential of
the 2HDM has been performed in
Refs.~\cite{Ivanov:2007de, Ferreira:2009wh}.
In the notation of Ref.~\cite{Ferreira:2009wh} they are:
the Higgs family symmetries $Z_2$, $U(1)$,
and $U(2)$;
and the (generalized) CP symmetries
CP1, CP2, and CP3.
Some steps have been taken to extend this analysis
into the Yukawa sector.
Ref.~\cite{Ferreira:2010ir} identifies all implementations of abelian
symmetries, showing that they allow for 34 distinct
matrix forms,
while Ref.~\cite{Ferreira:2010bm} shows that there is a single
implementation of a generalized CP symmetry
in both scalar potential and Yukawa couplings
which could be consistent with experiment.
There is no classification of all possible implementations
on non-abelian symmetries in both scalar and quark Yukawa sectors.
This is what we target here,
focusing on the simplest non-abelian group: $S_3$.

There has been recent interest in specific implementations
of $S_3$ in the 2HDM \cite{Kajiyama:2013sza, Johansen:2015nxa},
and many articles extending the 2HDM with extra
scalar and/or fermion fields --
see, for example, Refs.~\cite{Ma:2004pt, Ma:2013zca}.
In this article,
we provide a complete classification of all possible
implementation of $S_3$ in the 2HDM which are consistent with
non-vanishing quark masses and a CKM matrix which is not block-diagonal,
as required by experiment.

In Section~\ref{sec:S3} we introduce
the $S_3$ representation we use, discussing in detail
the features of the scalar potential implied by each
possible representation of the scalar fields,
the possible soft-breaking terms, and the
corresponding vacuum expectation values (vev).
In Section~\ref{sec:Yuk} we discuss all possible
extensions of $S_3$ into the Yukawa sector,
and in Section~\ref{sec:conclusions} we draw our conclusions.

\section{\label{sec:S3}The $S_3$ symmetry and the Higgs potential}

$S_3$ is the group of permutations of three objects
and it has 6 elements and three irreducible representations;
namely, two singlets
$\ \mathbf{1}, \ \mathbf{1^{\prime}}$,
and one doublet $\ \mathbf{2}$.
The multiplication rules are
\begin{eqnarray} \label{A4rules}
\mathbf{1} \otimes \textrm{any}
&=&
\textrm{any},
\nonumber\\
\mathbf{1^\prime} \otimes \mathbf{1^\prime}
&=&
\mathbf{1},
\nonumber\\
\mathbf{1^\prime} \otimes \mathbf{2}
&=&
\mathbf{2},
\\
\mathbf{2} \otimes \mathbf{2}
&=&
\mathbf{1} \oplus \mathbf{1^\prime} \oplus \mathbf{2}.
\nonumber
\end{eqnarray}
The representation $\mathbf{1}$ leaves the fields invariant,
while fields in $\mathbf{1^\prime}$ change their sign for odd permutations.
Consider a two Higgs doublet model (2HDM) with scalar fields
$\Phi_1$ and $\Phi_2$.

We will denote $\Phi \sim (\mathbf{1}, \mathbf{1}) $ when
both scalars are in the singlet representation of $S_3$.
In this case we obtain
the generic scalar potential of the 2HDM,
which may be written as \cite{hhg,ourreview}
\ba
V_H
&=&
m_{11}^2 |\Phi_1|^2
+ m_{22}^2 |\Phi_2|^2
- m_{12}^2\, \Phi_1^\dagger \Phi_2
- (m_{12}^2)^\ast\, \Phi_2^\dagger \Phi_1
\nonumber\\*[2mm]
&&
+\, \frac{\lambda_1}{2} |\Phi_1|^4
+ \frac{\lambda_2}{2} |\Phi_2|^4
+ \lambda_3 |\Phi_1|^2 |\Phi_2|^2
+ \lambda_4\, (\Phi_1^\dagger \Phi_2)\, (\Phi_2^\dagger \Phi_1)
\nonumber\\*[2mm]
&&
+\,
\left[
\frac{\lambda_5}{2} (\Phi_1^\dagger \Phi_2)^2
+ \left(\lambda_6 |\Phi_1|^2 + \lambda_7 |\Phi_2|^2\right)
(\Phi_1^\dagger \Phi_2)
+ \textrm{h.c.}
\right],
\label{VH}
\ea
where ``h.c.'' stands for the hermitian conjugate.
Hermiticity implies that all couplings are real,
except possibly $m_{12}^2$, $\lambda_5$, $\lambda_6$, and $\lambda_7$.
We will denote the vacuum expectation values (vevs) by $v_1/\sqrt{2}$
and $v_2/\sqrt{2}$,
but we will omit the $\sqrt{2}$ in the text
(though including it in actual calculations).
Depending on the parameters of the potential,
we may get any values for $(v_1, v_2)$,
including the inert cases $(v, 0)$ and $(0, v)$.

If we choose the $S_3$ representations as
$\Phi \sim (\mathbf{1}, \mathbf{1^\prime})$,
then all terms odd in $\Phi_2$ must vanish,
and we obtain the $Z_2$ symmetric potential:
$m_{12}^2=\lambda_6=\lambda_7=0$.
Usually, one includes also the terms in $m_{12}^2$,
which break the symmetry,
but only softly.
In this case,
if $\textrm{arg}(\lambda_5) = 2\, \textrm{arg}(m_{12}^2)$,
then the phases can be removed.
This is known as the real 2HDM and the scalar sector preserves CP.
One can have nonvanishing vevs $(v_1, v_2)$,
but the inert vevs $(v, 0)$ and $(0, v)$ are only possible if
$m_{12}^2 = 0$.
Alternatively,
if $\textrm{arg}(\lambda_5) \neq 2\, \textrm{arg}(m_{12}^2)$,
then the phases cannot be removed.
This is known as the complex 2HDM (C2HDM)
and the scalar sector violates CP.
In both models with softly broken $Z_2$ symmetry,
the conditions for a bounded from below potential are \cite{Deshpande:1977rw}
\be
\lambda_1 > 0,
\ \ \
\lambda_2 > 0,
\ \ \
\sqrt{\lambda_1 \lambda_2} > -\lambda_3,
\ \ \
\sqrt{\lambda_1 \lambda_2} > \left| \lambda_5 \right| - \lambda_3 - \lambda_4.
\label{bfb}
\ee

We denote collectively by $\Phi \sim \mathbf{s}$ the two possibilities just discussed:
$\Phi \sim (\mathbf{1}, \mathbf{1})$ and $\Phi \sim (\mathbf{1}, \mathbf{1^\prime})$.
Since the labels $1$ and $2$ are arbitrary,
the case $\Phi \sim (\mathbf{1^\prime}, \mathbf{1})$ is automatically included.
Moreover, the scalar potential remains the most general if one considers
$\Phi \sim (\mathbf{1^\prime}, \mathbf{1^\prime})$ \footnote{The Yukawa
couplings do not; but one can always reduce this to $\Phi \sim (\mathbf{1}, \mathbf{1})$.
For example,
if the right-handed fermions are in singlet representations of $S_3$,
one can change $\Phi \sim (\mathbf{1^\prime}, \mathbf{1^\prime})$ to
$\Phi \sim (\mathbf{1}, \mathbf{1})$,
by simultaneously changing the representation for the right-handed fermions
from $\mathbf{1}$ to $\mathbf{1^\prime}$ and vice-versa.}.

In order to describe how $S_3$ acts on doublets we must choose a
representation for the group.
A suitable real basis for the doublet representation of $S_3$ is given by
\cite{david}
\begin{equation}
a =
\left(
\begin{array}{cc}
1 & 0 \\
0 & -1
\end{array}
\right),
\hspace{10mm}
b =
\left(
\begin{array}{cc}
- \tfrac{1}{2} & - \tfrac{\sqrt{3}}{2}\\
\tfrac{\sqrt{3}}{2} & - \tfrac{1}{2}
\end{array}
\right).
\label{a&b}
\end{equation}
These matrices satisfy $a^2=b^3=(ba)^2= 1$,
showing that they indeed generate the group $S_3$.
In $S_3$, with the basis of Eq.~\eqref{a&b}, the product of two doublets,
$x=(x_1, x_2)^\intercal$ and $y=(y_1, y_2)^\intercal$, gives
\begin{eqnarray}
&&(x \otimes y)_{\mathbf{1}} =
x_1 y_1 + x_2 y_2,
\nonumber\\
&&(x \otimes y)_{\mathbf{1^\prime}} =
x_1 y_2 - x_2 y_1,
\nonumber\\
&&(x \otimes y)_{\mathbf{2}} =
(x_2 y_2 - x_1 y_1, x_1 y_2 + x_2 y_1)^\intercal.
\label{22}
\end{eqnarray}
Similarly, the product of the doublet $x$ with the singlet $y^\prime$
of $\mathbf{1^\prime}$ gives
\be
(x \otimes y^\prime) =
(-x_2 y^\prime, x_1 y^\prime)^\intercal.
\label{21prime}
\ee
%

Let us consider two scalars $\Phi = (\vp_1, \vp_2)^\intercal$ which transform
as a doublet under the real basis of Eq.~\eqref{a&b}.
According to Eq.~\eqref{22},
the relevant combinations of $\vp^\dagger_i \vp_j$ are
\begin{eqnarray}
&&
|\vp_2|^2 + |\vp_1|^2,
\nonumber\\
&&
\vp_1^\dagger \vp_2 - \vp_2^\dagger \vp_1,
\nonumber\\
&&
(|\vp_2|^2 - |\vp_1|^2,
\vp_1^\dagger \vp_2 + \vp_2^\dagger \vp_1)^\intercal,
\label{real_quadratic}
\end{eqnarray}
transforming, respectively, as $\mathbf{1}$,
$\mathbf{1^\prime}$, and $\mathbf{2}$.
Thus, there is only one quadratic term in the potential,
proportional to the first line of Eq.~\eqref{real_quadratic}.
The quartic terms come from squaring the first two lines and from
the singlet combination of two doublets of the third line.
Thus, the most general potential of a doublet of $S_3$,
consistent with the real representation of Eq.~\eqref{a&b} is:
\ba
V_R
&=&
\mu \left( |\vp_2|^2 + |\vp_1|^2 \right)
+
d_1 \left( |\vp_2|^2 + |\vp_1|^2 \right)^2
+
d_2  \left( \vp_1^\dagger \vp_2 - \vp_2^\dagger \vp_1 \right)^2
\nonumber\\
& &
+\, d_3 \left[
\left( |\vp_2|^2 - |\vp_1|^2 \right)^2
+
\left( \vp_1^\dagger \vp_2 + \vp_2^\dagger \vp_1 \right)^2
\right].
\label{VH_real}
\ea
This coincides with the generic potential in Eq.~\eqref{VH},
subject to the conditions
\be
m_{11}^2 = m_{22}^2,\ \
m_{12}^2 = 0,\ \
\lambda_1 = \lambda_2,\ \
\lambda_5 = \lambda_1 - \lambda_3 - \lambda_4,
\label{CP3}
\ee
identified in Table I of Ref.~\cite{Ferreira:2009wh} as the
CP3 model.

We have studied how the scalars transform as doublets under $S_3$
using the real representation in Eq.~\eqref{a&b}.
But other representations are possible.
Imagine that the scalars $\Phi$ transform under a Higgs family symmetry $S$ as
\be
\Phi \ra \Phi^S = S\, \Phi.
\ee
If one chooses a different basis, with new scalars $\Phi^\prime$
given by
\be
\Phi^\prime = U\, \Phi,
\label{change_U}
\ee
then the specific form of the Higgs family symmetry in the new basis
\be
\Phi^\prime \ra \Phi^{\prime S} = S^\prime\, \Phi^\prime
\ee
gets altered into
\be
S^\prime = U\, S\, U^\dagger.
\ee
As stressed by Ma \cite{Ma:2004pt},
it is easier to work with a complex representation for $S_3$,
obtained with
\be
U =
\frac{1}{\sqrt{2}}
\left(
\begin{array}{cc}
1 & i\\
1 & -i
\end{array}
\right)
\ \ \Longrightarrow\ \
U^\dagger =
\frac{1}{\sqrt{2}}
\left(
\begin{array}{cc}
1 & 1\\
-i & i
\end{array}
\right).
\label{matrix_U}
\ee
Applying this to the matrices in Eq.~\eqref{a&b},
we obtain
\begin{equation}
a_C =
\left(
\begin{array}{cc}
0 & 1 \\
1 & 0
\end{array}
\right),
\hspace{10mm}
b_C =
\left(
\begin{array}{cc}
\omega & 0\\
0 & \omega^2
\end{array}
\right),
\label{a&b_complex}
\end{equation}
where $a_C$ interchanges the two fields and $\omega = e^{2 i \pi/3}$ ($\omega^3=1$).
We denote the fields in the complex basis by
$\Phi^\prime =\mathbf{\phi}=(\phi_1, \phi_2)^\intercal$.
In $S_3$, with the basis of Eq.~\eqref{a&b_complex}, the product of two doublets,
$x=(x_1, x_2)^\intercal$ and $y=(y_1, y_2)^\intercal$, gives~\cite{Ma:2004pt}
\begin{eqnarray}
&&(x \otimes y)_{\mathbf{1}} =
x_1 y_2 + x_2 y_1,
\nonumber\\
&&(x \otimes y)_{\mathbf{1^\prime}} =
x_1 y_2 - x_2 y_1,
\nonumber\\
&&(x \otimes y)_{\mathbf{2}} =
(x_2 y_2, x_1 y_1)^\intercal.
\label{22_complex}
\end{eqnarray}
Comparing Eqs.~\eqref{22} and \eqref{22_complex} we see the great advantage
of the complex representation; the product of two doublets becomes diagonal,
as does the product of three doublets, which becomes simply $111+222$.
In some cases, such as that discussed in subsection~\ref{subsec:ex1} below,
the Yukawa couplings of the quarks are off-diagonal when
using the real representation.
However, one then has to diagonalize the mass matrices,
realizing that this diagonalization is performed by the same unitary
transformations in the up and down sectors and, thus,
that the CKM matrix coincides with the unit matrix.
The physics is easier to grasp in the complex representation,
where some Yukawa couplings become diagonal directly.
In the complex representation,
the product of the doublet $x$ with the singlet $y^\prime$
of $\mathbf{1^\prime}$ gives
\be
(x \otimes y^\prime) =
(x_1 y^\prime, -x_2 y^\prime)^\intercal.
\label{21prime_complex}
\ee

When considering the hermitian conjugated fields there is a complication arising
in the complex representation.
If $(\phi_1, \phi_2) \sim \mathbf{2}$,
then in the real representation $(\phi_1^\dagger, \phi_2^\dagger) \sim \mathbf{2}$,
but in the complex representation one has instead
$(\phi_2^\dagger, \phi_1^\dagger) \sim \mathbf{2}$ \cite{Ma:2004pt}.
Therefore,
considering two scalars $\mathbf{\phi}=(\phi_1, \phi_2)^\intercal$ which transform
as a doublet under the complex basis of Eq.~\eqref{a&b_complex},
the relevant combinations of $\phi^\dagger_i \phi_j$ according
to Eq.~\eqref{22_complex} are
\begin{eqnarray}
&&
|\phi_2|^2 + |\phi_1|^2,
\nonumber\\
&&
|\phi_2|^2 - |\phi_1|^2,
\nonumber\\
&&
(\phi^\dagger_1 \phi_2, \phi^\dagger_2 \phi_1)^\intercal,
\label{complex_quadratic}
\end{eqnarray}
transforming, respectively, as $\mathbf{1}$,
$\mathbf{1^\prime}$, and $\mathbf{2}$.
Thus, there is only one quadratic term in the potential,
proportional to the first line of Eq.~\eqref{complex_quadratic}.
The quartic terms come from squaring the first two lines and from
the singlet combination of two doublets of the third line.
Thus, the most general potential of a doublet of $S_3$,
consistent with the complex representation of Eq.~\eqref{a&b_complex} is
\cite{Ma:2013zca}:
\ba
V_C
&=&
\mu_1^2 \left( |\phi_2|^2 + |\phi_1|^2 \right)
+
\tfrac{1}{2} \ell_1 \left( |\phi_2|^2 + |\phi_1|^2 \right)^2
+
\tfrac{1}{2} \ell_2  \left( |\phi_2|^2 - |\phi_1|^2 \right)^2
\nonumber\\
& &
+\, \ell_3 ( \phi_1^\dagger \phi_2 ) (\phi_2^\dagger \phi_1).
\label{VH_complex}
\ea
This is the same as Eq.~\eqref{VH_real},
through the transformation in Eqs.~\eqref{change_U} and \eqref{matrix_U},
with $\mu_1^2=\mu$, $\ell_1 = 2 d_1$, $\ell_2 = - 2 d_2$, and $\ell_3 = 4 d_3$.
The potential obtained coincides with the generic potential in Eq.~\eqref{VH},
subject to the conditions
\be
m_{11}^2 = m_{22}^2,\ \
m_{12}^2 = 0,\ \
\lambda_1 = \lambda_2,\ \
\lambda_5 = 0,
\label{CP3_reduced}
\ee
which may seem not to coincide with those in Eq.~\eqref{CP3}.
However, they are the same conditions, but seen in different basis.
This had already been pointed out in Eqs.~(104)-(106) of Ref.~\cite{Ferreira:2009wh},
where it is shown that the CP3 model can also be obtained by imposing
both $U(1)$ and the so-called $\Pi_2$ symmetry $\phi_1 \leftrightarrow \phi_2$
in the same basis.
Ma and Melic \cite{Ma:2013zca} also include in the potential a term which
breaks $S_3$ softly,
while preserving the $\phi_1 \leftrightarrow \phi_2$ symmetry:
\be
V_\textrm{soft} = - \mu_2^2
( \phi_1^\dagger \phi_2  + \phi_2^\dagger \phi_1).
\ee
This term is needed since otherwise there would be a massless pseudoscalar.
But, including it precludes the inert vevs $(v,0)$ and $(0,v)$.

We will now consider the potential $V = V_C + V_\textrm{soft}$.
In terms of the new parameters, the bounded from below conditions
in Eq.~\eqref{bfb} read
\be
\ell_1 + \ell_2 > 0,
\ \ \
\ell_1 > 0,
\ \ \
2 \ell_1 + \ell_3 > 0.
\label{bfb_V}
\ee
A convenient way to write the stationarity conditions is
\ba
0 = v_2 \frac{\partial V}{\partial v_1}
- v_1 \frac{\partial V}{\partial v_2}
&=&
(v_1^2 - v_2^2)
\left[
\mu_2^2 + \tfrac{1}{2}  (2 \ell_2 - \ell_3) v_1 v_2
\right],
\nonumber\\
0 = v_1 \frac{\partial V}{\partial v_1}
- v_2 \frac{\partial V}{\partial v_2}
&=&
(v_1^2 - v_2^2)
\left[
\mu_1^2 + \tfrac{1}{2} (\ell_1 + \ell_2) v^2
\right],
\label{min}
\ea
where $v = \sqrt{v_1^2 + v_2^2} = 246\ \textrm{GeV}$.
Ma and Melic \cite{Ma:2013zca} consider the vev $(v,v)$,
for which $\partial V/\partial v_1$ yields
\be
\mu_1^2 = \mu_2^2 - \tfrac{1}{4} (2 \ell_1 + \ell_3) v^2
\ee
and the scalar masses
\ba
m_{H^\pm}^2
&=&
2 \mu_2^2 - \tfrac{1}{2} \ell_3 v^2,
\nonumber\\
m_A^2
&=&
2 \mu_2^2,
\nonumber\\
m_h^2
&=&
\tfrac{1}{2} (2 \ell_1 + \ell_3) v^2,
\nonumber\\
m_H^2
&=&
2 \mu_2^2 + \tfrac{1}{2} (2 \ell_2 - \ell_3) v^2,
\label{masses_vv}
\ea
for the charged scalars ($H^\pm$),
the pseudoscalar ($A$),
the light ($h$) and the heavy ($H$) CP even
scalars, respectively.

We consider also the vev\footnote{Recall that when the text refers to
$(v_1, v_2)$ what is really meant is $\langle \phi_k^0 \rangle = v_k/\sqrt{2}$
($k=1,2$), and that this is what we use in the equations.}
$(v_1, v_2)$ with $v_1 \neq v_2$.
Then, $\mu_1^2$ and $\mu_2^2$ are found from Eqs.~\eqref{min}.
We parametrize the fields as
\ba
\phi_1
&=&
\left(
\begin{array}{c}
c_\beta G^+ - s_\beta H^+\\*[1mm]
\frac{1}{\sqrt{2}}
\left[
v_1 + \rho_1
+ i \left(
c_\beta G^0 - s_\beta A
\right)
\right]
\end{array}
\right),
\nonumber\\
\phi_2
&=&
\left(
\begin{array}{c}
s_\beta G^+ + c_\beta H^+\\*[1mm]
\frac{1}{\sqrt{2}}
\left[
v_2 + \rho_2
+ i \left(
s_\beta G^0 + c_\beta A
\right)
\right]
\end{array}
\right),
\ea
where $G^0$ and $G^+$ are the would-be Goldstone bosons,
\be
v_1 = v\, c_\beta, \ \ v_2 = v\, s_\beta,
\ee
and,
thenceforth,
$c_\theta$ ($s_\theta$, $t_\theta$) represent the cosine
(sine, tangent) of whatever angle $\theta$ is the subindex.
The charged scalar and pseudoscalar masses become
\ba
m_{H^\pm}^2
&=&
- \ell_2  v^2,
\label{mC}
\\
m_A^2
&=&
- \tfrac{1}{2} (2 \ell_2 - \ell_3) v^2,
\label{mA}
\ea
while the CP even scalar mass matrix is
\be
M_n =
\left(
\begin{array}{cc}
\ell_1 v_1^2 + \tfrac{1}{2} \ell_3 v_2^2 + \ell_2 (v_1^2 - v_2^2)
& \tfrac{1}{2} (2 \ell_1 + \ell_3) v_1 v_2 \\*[2mm]
\tfrac{1}{2} (2 \ell_1 + \ell_3) v_1 v_2
& \ell_1 v_2^2 + \tfrac{1}{2} \ell_3 v_1^2 - \ell_2 (v_1^2 - v_2^2)
\end{array}
\right).
\ee
Its trace and determinant are
\ba
m_h^2 + m_H^2 = \textrm{Tr}\left(M_n\right)
&=&
\tfrac{1}{2} (2 \ell_1 + \ell_3) v^2
\label{tr}
\\
m_h^2\,  m_H^2 = \textrm{Det}\left(M_n\right)
&=&
- \tfrac{1}{2} (\ell_1 + \ell_2) (2 \ell_2 - \ell_3) (v_1^2 - v_2^2)^2.
\label{det}
\ea
Notice that, for negative $\ell_2$ such that
$-\ell_1 < \ell_2 < \ell_3/2$, the requirement that squared masses are
positive can be made consistent with the bounded from below conditions in
Eq.~\eqref{bfb_V}.
Moreover,
Eqs.~\eqref{masses_vv} cannot be obtained as the limit
$v_1, v_2 \rightarrow v/\sqrt{2}$ of Eqs.~\eqref{mC}-\eqref{det} --
Eq.~\eqref{det} would lead to a massless scalar;
this is a singular case which must be treated separately.
Eqs.~\eqref{mC}, \eqref{mA}, and \eqref{tr} can be inverted, to yield
$\ell_1$, $\ell_2$, and $\ell_3$ in terms of the physical masses:
\ba
\ell_1 v^2 &=&
m_h^2 + m_H^2 + m_{H^\pm}^2 - m_A^2,
\nonumber\\
\ell_2 v^2 &=&
- m_{H^\pm}^2,
\nonumber\\
\ell_3 v^2 &=&
2( m_A^2 - m_{H^\pm}^2).
\ea
Substituting into Eq.~\eqref{det},
we obtain
\be
\cos^2{(2 \beta)}
= \frac{m_h^2\ m_H^2}{m_A^2 (m_h^2 + m_H^2 - m_A^2)}.
\label{cos2beta}
\ee

The diagonalization of $M_n$ is performed through the transformation
\be
\left(
\begin{array}{c}
\textrm{Re}\, \phi_1^0\\*[1mm]
\textrm{Re}\, \phi_2^0
\end{array}
\right)
=
\left(
\begin{array}{c}
\rho_1 \\*[1mm]
\rho_2
\end{array}
\right)
=
\left(
\begin{array}{cc}
\cos{\alpha} & -\sin{\alpha}\\*[1mm]
\sin{\alpha} & \cos{\alpha}
\end{array}
\right)
\,
\left(
\begin{array}{c}
H\\*[1mm]
h
\end{array}
\right).
\ee
We find
\be
\tan{(2 \alpha)}
=
\frac{2 \ell_1 + \ell_3}{2 \ell_1 + 4 \ell_2 - \ell_3} \frac{2 v_1 v_2}{v_1^2 - v_2^2}
=
\frac{m_h^2 + m_H^2}{m_h^2 + m_H^2 - 2 m_A^2} \tan{(2 \beta)}.
\label{tan2alpha}
\ee
Using $\tan^2{(2 \beta)} = 1/\cos^2{(2 \beta)} - 1$ and Eq.~\eqref{cos2beta},
one can write $\tan{(2 \alpha)}$ in terms of masses alone.
In the real 2HDM,
the coupling of the lighter scalar to two vector bosons is given by
$g_{hVV} = g_{hVV}^\textrm{sm} \sin{(\beta-\alpha)}$,
where $g_{hVV}^\textrm{sm}$ is the SM coupling.
Measurements of these couplings at LHC constrain $\beta-\alpha$
to lie very close to $\pi/2$ -- the so-called alignment limit.
Through Eqs.~\eqref{cos2beta} and \eqref{tan2alpha},
this places tight constraints on the masses of $A$ and $H$.
In the exact alignment limit, we would be forced into $m_A=0$.
Thus, in this case the pseudoscalar must be light,
running into constraints from both LEP and LHC.
A full phenomenological analysis of this case is beyond the scope of this work.

We now turn to the $S_3$ potential with the most general real
soft violations of $S_3$:
\ba
V
&=&
\mu_1^2 \left( |\phi_2|^2 + |\phi_1|^2 \right)
- \mu_2^2
( \phi_1^\dagger \phi_2  + \phi_2^\dagger \phi_1)
- \mu_3^2 \left( |\phi_2|^2 - |\phi_1|^2 \right)
\nonumber\\
&&
+\,
\tfrac{1}{2} \ell_1 \left( |\phi_2|^2 + |\phi_1|^2 \right)^2
+
\tfrac{1}{2} \ell_2  \left( |\phi_2|^2 - |\phi_1|^2 \right)^2
+
\ell_3 ( \phi_1^\dagger \phi_2 ) (\phi_2^\dagger \phi_1).
\label{V_gensoft}
\ea
The stationarity conditions become
\ba
0 = \frac{\partial V}{\partial v_1}
&=&
(\mu_1^2 + \mu_3^2) v_1 - \mu_2^2 v_2
+ \tfrac{v_1}{2}
\left(
\ell_1 v^2 + \ell_2 (v_1^2 - v_2^2) + \ell_3 v_2^2
\right),
\nonumber\\
0 = \frac{\partial V}{\partial v_2}
&=&
(\mu_1^2 - \mu_3^2) v_2 - \mu_2^2 v_1
+ \tfrac{v_2}{2}
\left(
\ell_1 v^2 - \ell_2 (v_1^2 - v_2^2) + \ell_3 v_1^2
\right),
\label{stationary_gensoft}
\ea
and $v_1=v_2$ is no longer a minimum.
Although it is technically natural to have $\mu_3^2$ small
(since in the $\mu_3^2=0$ limit the potential gets an extra
$Z_2$ symmetry),
in which case $v_1 \sim v_2$,
we will allow any value for $\mu_3^2$.
Repeating the previous steps,
we find
\ba
m_{H^\pm}^2
&=&
- \ell_2  v^2 - 2 \mu_3^2 \sec{(2 \beta)},
\label{mC_3}
\\
m_A^2
&=&
- \tfrac{1}{2} \left[(2 \ell_2 - \ell_3) v^2
+ 4 \mu_3^2 \sec{(2 \beta)}
\right],
\label{mA_3}
\\
T \equiv m_h^2 + m_H^2
&=&
\tfrac{1}{2} \left[(2 \ell_1 + \ell_3) v^2
- 4 \mu_3^2 \sec{(2 \beta)}
\right],
\label{tr_3}
\\
D \equiv m_h^2\,  m_H^2
&=&
- \frac{v^2}{2}
\left[
(2 \ell_2 - \ell_3) \cos{(2 \beta)}
\left(
(\ell_1 + \ell_2) v^2 \cos{(2 \beta)} + 2 \mu_3^2
\right)
+ 2 (2 \ell_1 + \ell_3) \mu_3^2 \sec{(2 \beta)}
\right].
\label{det_3}
\ea
Hence,
\ba
\ell_1 v^2 &=&
m_h^2 + m_H^2 + m_{H^\pm}^2 - m_A^2 + 2 \mu_3^2 \sec{(2 \beta)},
\nonumber\\
\ell_2 v^2 &=&
- m_{H^\pm}^2 - 2 \mu_3^2 \sec{(2 \beta)},
\nonumber\\
\ell_3 v^2 &=&
2( m_A^2 - m_{H^\pm}^2).
\ea
These results reproduce the corresponding ones in the
$\mu_3^2 \ra 0$ limit.
Substituting into Eq.~\eqref{det_3} yields
\ba
m_h^2\,  m_H^2
&=&
m_A^2 (m_h^2 + m_H^2 - m_A^2) \cos^2{(2 \beta)}
\nonumber\\
&&
- 2 \mu_3^2 (m_h^2 + m_H^2) \cos{(2 \beta)} \tan^2{(2 \beta)}
- 4 (\mu_3^2)^2 \tan^2{(2 \beta)},
\label{det_4}
\ea
providing a complicated relation between $\mu_3$ and $\beta$.
Also
\be
\tan{(2 \alpha)}
=
\frac{m_h^2 + m_H^2 + 4 \mu_3^2 \sec{(2 \beta)} }{
m_h^2 + m_H^2 - 2 m_A^2} \tan{(2 \beta)}.
\label{tan2alpha_3}
\ee
Thus,
for the potential of Eq.~\eqref{V_gensoft},
the independent parameters may be taken
as $v^2$, $\beta$, $m_h^2$, $m_H^2$, $m_A^2$, and $m_{H^\pm}^2$;
$\mu_3^2$ is found from Eq.~\eqref{det_4} and $\alpha$ from Eq.~\eqref{tan2alpha_3}.
Combining Eqs.~\eqref{det_4} and \eqref{tan2alpha_3}, we find
\be
\frac{D}{c_{2 \beta}^2}
=
m_A^2 (T - m_A^2) + \frac{T^2}{4} t_{2 \beta}^2 -
\left(
\frac{T}{2} - m_A^2
\right)^2 t_{2 \alpha}^2.
\label{new}
\ee
Even in the exact alignment limit ($\beta = \alpha + \pi/2$),
this reduces to $m_A^2(T-m_A^2)=D$ which has two solutions:
$m_A^2 =m_h^2$; and the definitely allowed $m_A^2 = m_H^2$,
consistent with the decoupling limit.

So far, we have considered only real parameters and vevs in the scalar sector.
We now turn briefly to the possibility that they are complex.
Since the $S_3$ symmetry forces $\lambda_5= \lambda_6 = \lambda_7 =0$,
one can only introduce phases via soft breaking,
by giving a phase $\chi$ to $\mu_2^2$ and
changing in Eq.~\eqref{V_gensoft} the term
\be
- \mu_2^2
( \phi_1^\dagger \phi_2  + \phi_2^\dagger \phi_1)
\ra
- \mu_2^2
( e^{i \chi} \phi_1^\dagger \phi_2  + e^{-i \chi}\phi_2^\dagger \phi_1).
\ee
The most general vev may be written as $v_1$, $v_2 e^{i \delta}$.
Minimization with respect to $\delta$,
\be
0 = \frac{\partial V}{\partial \delta}
=
\mu_2^2 v_1 v_2 \sin{(\delta + \chi)},
\ee
forces\footnote{The other possibilities --
$\mu_2^2 = 0$, or inert vacua $v_1=0$ or $v_2=0$
--
will not be pursued here.}
$\delta = - \chi$,
while the other minimization conditions,
in Eqs.~\eqref{stationary_gensoft} remain the same.
Thus, we can only have complex vevs with a complex
soft-breaking term.

\section{\label{sec:Yuk}Yukawa couplings}

We now turn to the Yukawa couplings,
following closely the notation of Ref.~\cite{Nebot:2015wsa}:
\be
- {\cal L}_Y = \bar{q}_L (\Gamma_1 \Phi_1 + \Gamma_2 \Phi_2) n_R
+
\bar{q}_L (\Delta_1 \tilde{\Phi}_1 + \Delta_2 \tilde{\Phi}_2) p_R + \textrm{h.c.},
\ee
where  $q_L = (p_L, n_L)^\intercal$  is a vector in the 3-dimensional
family space of left-handed doublets,
and $n_R$ ($p_R$) is a vector in the 3-dimensional right-handed
space of charge $-1/3$ ($+2/3$) quarks.
The complex $3 \times 3$ matrices
$\Gamma_1$, $\Gamma_2$, $\Delta_1$, and $\Delta_2$ contain the Yukawa couplings.
In general, these matrices are not diagonal.
We denote by $U_\alpha$ ($\alpha = d_L, d_R, u_L, u_R$) the matrices taking
the quarks into the mass basis:
\ba
&&
n_R = U_{d_R}\, d_R,
\hspace{5ex}
p_R = U_{u_R}\, u_R,
\nonumber\\
&&
\bar{q}_L = (\bar{p}_L,\ \bar{n}_L) =
(\bar{u}_L\, U_{u_L}^\dagger,\ \bar{d}_L\, U_{d_L}^\dagger)
=
(\bar{u}_L\, V,\ \bar{d}_L)  U_{d_L}^\dagger,
\ea
where $V = U_{u_L}^\dagger U_{d_L}$ is the CKM matrix.
The mass matrices become
\ba
\textrm{diag} (m_d, m_s, m_b)
=
D_d
& = &
\frac{1}{\sqrt{2}} U_{d_L}^\dagger
\left[
v_1\, \Gamma_1 + v_2 \Gamma_2
\right]
U_{d_R},
\nonumber\\
\textrm{diag} (m_u, m_c, m_t )
=
D_u
& = &
\frac{1}{\sqrt{2}} U_{u_L}^\dagger
\left[
v_1\, \Delta_1 + v_2 \Delta_2 \right]
U_{u_R},
\ea
while eventual flavour changing neutral scalar interactions
are controlled by
\ba
N_d
& = &
\frac{1}{\sqrt{2}} U_{d_L}^\dagger
\left[
- v_2\, \Gamma_1 + v_1 \Gamma_2
\right]
U_{d_R},
\nonumber\\
N_u
& = &
\frac{1}{\sqrt{2}} U_{u_L}^\dagger
\left[
- v_2\, \Delta_1 + v_1 \Delta_2 \right]
U_{u_R}.
\ea
In this section, we will absorb the $1/\sqrt{2}$ in the definitions of the
parameters in the Yukawa matrices, and work with
\be
Y_d = v_1\, \Gamma_1 + v_2 \Gamma_2,
\ \ \ \
Y_u = v_1\, \Delta_1 + v_2 \Delta_2,
\ee
and the hermitian matrices
\ba
H_d = Y_d Y_d^\dagger
&=&
U_{d_L} \, \textrm{diag} (m_d^2, m_s^2, m_b^2)\, U_{d_L}^\dagger,
\nonumber\\
H_u = Y_u Y_u^\dagger
&=&
U_{u_L} \, \textrm{diag} (m_u^2, m_c^2, m_t^2)\, U_{u_L}^\dagger.
\label{HdHu}
\ea
These matrices are diagonalized respectively by the matrices
$U_{d_L}$ and $U_{u_L}$,
which must be found when fitting for the CKM matrix.
This step is not needed for CP violation,
which may be found directly from \cite{Ja}
\begin{equation}
J = \textrm{Det} (H_d H_u - H_u H_d),
\label{J}
\end{equation}
in a basis independent fashion.

Henceforth, we use the complex representation.
We recall that, for the corresponding entry of the Yukawa coupling matrix to
be non-vanishing, the Yukawa Lagrangian must be in the invariant singlet representation
$\mathbf{1}$.
We start by assuming that the two scalar fields are in singlet representations,
which we denoted before by $\Phi \sim \mathbf{s}$.
Thus,
the product of left-handed ($q_L$) and right-handed
(charge $+2/3$, $p_R$, or charge $-1/3$, $n_R$)
fermions must also be in a singlet representation.
This can be achieved by two doublets or by two singlets.
Since permutations of the three fields in each sector do not lead to new
structures for the Yukawa matrices,
we will denote by $f \sim \mathbf{s}$
(where $f = q_L, p_R, n_R$)
the following independent possibilities for the fields in each of the
three generations:
\begin{eqnarray}
(\mathbf{1}, \mathbf{1}, \mathbf{1}),
& \quad &
(\mathbf{1}, \mathbf{1}, \mathbf{1^\prime}),
\nonumber\\
(\mathbf{1}, \mathbf{1^\prime}, \mathbf{1^{\prime}}),
& \quad &
(\mathbf{1^\prime}, \mathbf{1^\prime}, \mathbf{1^\prime}).
\end{eqnarray}
Similarly,
we will denote by $f \sim \mathbf{d}$
the following independent possibilities for the fields in each of the
three generations:
\be
(\mathbf{2}, \mathbf{1}),
 \quad
(\mathbf{2}, \mathbf{1^\prime}).
\ee

Let us now consider the possibility that the two scalar fields are in
the doublet representation of $S_3$,
which we denoted before by $\Phi \sim \mathbf{2}$.
Thus,
the product of left-handed ($q_L$) and right-handed
(charge $+2/3$, $p_R$, or charge $-1/3$, $n_R$)
fermions must also be in a doublet representation.
But this can be achieved by having
both quark fields in doublets,
or by having one in a doublet and another in a singlet.
The possibilities for the various representations
are listed in Table~\ref{tabS3:rep}.
\begin{table}[h!]
\begin{center}
\begin{tabular}{|ccccccccc|}
\hline
& $\Phi$ & & $\bar{q}_L$ & & $n_R$ & & $p_R$ & \\
\hline
& $\mathbf{s}$ & & $\mathbf{s}$  & &
$\mathbf{s}$  & &
$\mathbf{s}$ & \\
& $\mathbf{s}$ & & $\mathbf{d}$  & &
$\mathbf{d}$  & &
$\mathbf{d}$ & \\
\hline
& $\mathbf{2}$ & & $\mathbf{d}$  & &
$\mathbf{d}$  & &
$\mathbf{d}$ & \\
& $\mathbf{2}$ & & $\mathbf{d}$  & &
$\mathbf{s}$  & &
$\mathbf{d}$  & \\
& $\mathbf{2}$ & & $\mathbf{d}$  & &
$\mathbf{d}$  & &
$\mathbf{s}$  & \\
& $\mathbf{2}$ & & $\mathbf{d}$  & &
$\mathbf{s}$ & &
$\mathbf{s}$ & \\
& $\mathbf{2}$ & & $\mathbf{s}$  & &
$\mathbf{d}$  & &
$\mathbf{d}$ & \\
\hline
\end{tabular}
\end{center}
\caption{Possible representations for the scalar fields ($\Phi$),
the left-handed quark $SU(2)_L$ doublets ($q_L$),
and the right-handed quark $SU(2)_L$ singlets ($p_R$ and $n_R$).
\label{tabS3:rep}}
\end{table}
The first case (with all fields in singlets),
reduces to the analysis of $Z_2$ and
has been discussed in detail in Ref.~\cite{Ferreira:2010ir},
where are possible implementations with abelian groups are presented.

\subsection{\label{subsec:ex1}Example 1: $\Phi$ in singlet; fermions in doublets}

As a specific example,
let us consider the possibility
\be
\Phi \sim (\mathbf{1}, \mathbf{1^\prime}), \ \
\bar{q}_L \sim (\mathbf{2}, \mathbf{1}), \ \
n_R \sim (\mathbf{2}, \mathbf{1}), \ \
p_R \sim (\mathbf{2}, \mathbf{1}).
\ee
%
Because $\Phi$ is in a singlet,
the product of fermions (left and right) must also be in a singlet.
Using $\bar{q}_L$ in the doublet,
the product of two doublets $\bar{q}_L n_R$ in $\mathbf{2} \otimes \mathbf{2}$ is
\be
\left.
\left(
\begin{array}{c}
\bar{q}_{L 1}\\
\bar{q}_{L 2}
\end{array}
\right)
\otimes
\left(
\begin{array}{c}
n_{R 1}\\
n_{R 2}
\end{array}
\right)
\right|_{\mathbf{1}, \mathbf{1^\prime}}
=
\bar{q}_{L 1} n_{R 2}
\pm
\bar{q}_{L 2} n_{R 1}.
\ee
Thus,
the products with the scalars into a singlet
are\footnote{We will often show the product of fields in the
order that highlights the $S_3$ properties, even if it
violates the correct $SU(2)_L$ order of
$\bar{q}_L \Phi f_R$.}
\ba
&&
\Phi_1
\left(
\bar{q}_{L 1} n_{R 2} + \bar{q}_{L 2} n_{R 1}
\right),
\label{for_a_1}
\\
&&
\Phi_2
\left(
\bar{q}_{L 1} n_{R 2} - \bar{q}_{L 2} n_{R 1}
\right).
\label{for_b_1}
\ea
The remaining non vanishing term comes from
\be
\Phi_1 \bar{q}_{L 3} n_{R 3}.
\label{for_c_1}
\ee
Multiplying Eqs.~\eqref{for_a_1}, \eqref{for_b_1},
and \eqref{for_c_1} by complex coefficients
$a$, $b$, and $c$, respectively,
we find
\be
Y_d
=
\left[
\begin{array}{ccc}
0 & a v_1 + b v_2 & 0\\
a v_1 - b v_2 & 0 & 0\\
0 & 0 & c v_1
\end{array}
\right]
=
\left[
\begin{array}{ccc}
0 & 1 & 0\\
1 & 0 & 0\\
0 & 0 & 1
\end{array}
\right]
\left[
\begin{array}{ccc}
a v_1 - b v_2 & 0 & 0\\
0 & a v_1 + b v_2 & 0\\
0 & 0 & c v_1
\end{array}
\right]
.
\label{Yd_1222}
\ee
This corresponds to the charged lepton sector of the model
discussed in Refs.~\cite{Kajiyama:2013sza, Johansen:2015nxa},
where neutrino mixing was not considered.
We notice that the matrix in Eq.~\eqref{Yd_1222} is block diagonal.
For the same reason,
so would be the corresponding matrix for the up quarks,
leading to a block diagonal CKM matrix,
in contradiction with experiment.
Thus,
these $S_3$ assignments cannot be used for the
quarks\footnote{In Refs.~\cite{Kajiyama:2013sza, Johansen:2015nxa}
this problem is solved by keeping all quark fields in singlets;
moreover, with the same assignment
($\mathbf{1}$ or $\mathbf{1^\prime}$) for
all generations within each sector.
This reduces the quark sector to the usual $Z_2$ cases.}.
Because $\Phi$ is in a singlet
and only one generation of quarks in each sector is in a singlet,
this block-diagonal problem remains,
regardless of the assignments of such singlets
to $\mathbf{1}$ or $\mathbf{1^\prime}$.
Thus,
we conclude that the case on the second line of Table~\ref{tabS3:rep},
cannot be implemented in the quark sector.

\subsection{Example 2: doublets in all sectors}

We now turn to
\be
\Phi \sim \mathbf{2}, \ \
\bar{q}_L \sim (\mathbf{2}, \mathbf{1}), \ \
n_R \sim (\mathbf{2}, \mathbf{1}), \ \
p_R \sim (\mathbf{2}, \mathbf{1}).
\ee
Because $\Phi$ is in a doublet,
the product of fermions (left and right) must also be in a doublet.
Using $\bar{q}_L$ in the doublet,
the product of two doublets $\bar{q}_L n_R$ in $\mathbf{2} \otimes \mathbf{2}$ is
\be
\left.
\left(
\begin{array}{c}
\bar{q}_{L 1}\\
\bar{q}_{L 2}
\end{array}
\right)
\otimes
\left(
\begin{array}{c}
n_{R 1}\\
n_{R 2}
\end{array}
\right)
\right|_\mathbf{2}
=
\left(
\begin{array}{c}
\bar{q}_{L 2} n_{R 2}\\
\bar{q}_{L 1} n_{R 1}
\end{array}
\right).
\ee
The product with the scalar doublet into a singlet is
\be
\left.
\left(
\begin{array}{c}
\Phi_1\\
\Phi_2
\end{array}
\right)
\otimes
\left(
\begin{array}{c}
\bar{q}_{L 2}\, n_{R 2}\\
\bar{q}_{L 1}\, n_{R 1}
\end{array}
\right)
\right|_\mathbf{1}
=
\Phi_1 \bar{q}_{L 1} n_{R 1} + \Phi_2 \bar{q}_{L 2} n_{R 2},
\label{for_a}
\ee
as mentioned after Eq.~\eqref{22_complex}.
For a $n_{R 3}$ in a singlet,
we find
\be
\left.
\left(
\begin{array}{c}
\Phi_1\\
\Phi_2
\end{array}
\right)
\otimes
\left(
\begin{array}{c}
\bar{q}_{L 1}\\
\bar{q}_{L 2}
\end{array}
\right)
\right|_\mathbf{1}
\otimes
n_{R 3}
=
\Phi_1 \bar{q}_{L 2} n_{R 3} + \Phi_2 \bar{q}_{L 1} n_{R 3}.
\label{for_b}
\ee
Finally, for a $\bar{q}_{L 3}$ in a singlet,
we find
\be
\bar{q}_{L 3}
\otimes
\left.
\left(
\begin{array}{c}
\Phi_1\\
\Phi_2
\end{array}
\right)
\otimes
\left(
\begin{array}{c}
n_{R 1}\\
n_{R 2}
\end{array}
\right)
\right|_\mathbf{1}
=
\Phi_1 \bar{q}_{L 3} n_{R 2} + \Phi_2 \bar{q}_{L 3} n_{R 1}.
\label{for_c}
\ee
Since $\Phi$ is in the doublet,
$d_L$ and $n_R$ cannot be simultaneously in the singlet.
Thus,
multiplying Eqs.~\eqref{for_a}, \eqref{for_b},
and \eqref{for_c} by complex coefficients
$a$, $b$, and $c$, respectively,
we find
\be
Y_d
=
\left[
\begin{array}{ccc}
a v_1 & 0 & b v_2\\
0 & a v_2 & b v_1\\
c v_2 & c v_1 & 0
\end{array}
\right].
\label{Yd_222}
\ee

The matrix for the up quarks is easily found by noting that the doublet
$(\Phi_1, \Phi_2)^\intercal$ gets substituted by
$(\tilde{\Phi}_2, \tilde{\Phi}_1)^\intercal$,
corresponding to a $v_1 \leftrightarrow v_2^\ast$ change.
We find
\be
Y_u
=
\left[
\begin{array}{ccc}
x v_2^\ast & 0 & y v_1^\ast\\
0 & x v_1^\ast & y v_2^\ast\\
z v_1^\ast & z v_2^\ast & 0
\end{array}
\right],
\label{Yu_222}
\ee
with complex coefficients $x$, $y$, and $z$.
Besides $v$ and $\tan{\beta}$,
the matrices involve 6 complex coefficients,
which might conceivably fit the ten known data:
the 6 quark masses and the 4 parameters in the CKM matrix.
We have used Eqs.~\eqref{HdHu} to check that
we can generate all masses different and nonzero,
and Eq.~\eqref{J} to show that we can generate
a nonzero CP violating phase.
We note that $J \neq 0$ even if one takes the vevs to be real,
implying that this model does not coincide with the
CP3 model with quarks presented in Ref.~\cite{Ferreira:2010bm},
where a complex vev was needed in order to get a non-vanishing
$J$.
This a further illustration of a sometimes unappreciated point:
two symmetries which lead to the same scalar potential,
may lead to very different models when extended into the
Yukawa sector.
A complete analysis of each model is beyond the scope of this
work. We are interested here in mapping all possibilities
consistent with $S_3$ which do not lead necessarily
into vanishing and/or degenerate masses,
or to the absence of CP violation.

We must now turn to the possibility that some among
$\bar{q}_{L 3}$, $n_{L 3}$, and/or $p_{L 3}$
are in a $\mathbf{1^\prime}$.
Using $\eta_\alpha = +1$ when the corresponding field
is in $\mathbf{1}$,
and $\eta_\alpha = -1$ when the corresponding field
is in $\mathbf{1^\prime}$, we find
\ba
Y_d
&=&
\left[
\begin{array}{ccc}
a v_1 & 0 & \eta_d b v_2\\
0 & a v_2 & b v_1\\
\eta_q c v_2 & c v_1 & 0
\end{array}
\right],
\nonumber\\
Y_u
&=&
\left[
\begin{array}{ccc}
x v_2^\ast & 0 & \eta_u y v_1^\ast\\
0 & x v_1^\ast & y v_2^\ast\\
\eta_q z v_1^\ast & z v_2^\ast & 0
\end{array}
\right].
\ea
These are all forms consistent with the assignments on the
third line of Table~\ref{tabS3:rep}.

\subsection{Example 3: singlet only on right-handed sectors}

Let us consider
\be
\Phi \sim \mathbf{2}, \ \
\bar{q}_L \sim (\mathbf{2}, \mathbf{1}), \ \
n_R \sim \mathbf{s}, \ \
p_R \sim (\mathbf{2}, \mathbf{1}).
\ee
Because $\Phi$ is in a doublet,
the product of fermions (left and right) must also be in a doublet.
Since $\bar{q}_{L 3}$ and all $n_R$ are in a singlet,
the last line of the matrix $Y_d$ vanishes.
But this implies that the matrix $H_d$ only has non-vanishing
entries in the $(1,2)$ sector,
leading to a massless down quark.
Similarly,
\be
\Phi \sim \mathbf{2}, \ \
\bar{q}_L \sim (\mathbf{2}, \mathbf{1}), \ \
n_R \sim (\mathbf{2}, \mathbf{1}), \ \
p_R \sim \mathbf{s},
\ee
leads to a massless up quark.
A combination of both problems occurs in
\be
\Phi \sim \mathbf{2}, \ \
\bar{q}_L \sim (\mathbf{2}, \mathbf{1}), \ \
n_R \sim \mathbf{s}, \ \
p_R \sim \mathbf{s},
\ee
Thus,
under the conditions considered in this paper
of $S_3$ in a 2HDM,
these three cases are ruled out.

\subsection{Example 4: singlet only on left-handed sector}

The last case to be considered is
\be
\Phi \sim \mathbf{2}, \ \
\bar{q}_L \sim \mathbf{s}, \ \
n_R \sim (\mathbf{2}, \mathbf{1}), \ \
p_R \sim (\mathbf{2}, \mathbf{1}).
\ee
Because $\Phi$ is in a doublet,
the product of fermions (left and right) must also be in a doublet.
Since all $\bar{q}_L$ and $n_{R 3}$ are in a singlet,
the last column of the matrix $Y_d$ vanishes.
But this implies that the matrix $Y_d^\dagger Y_d$
(which has the same eigenvalues as $H_d$)
only has non-vanishing entries in the $(1,2)$ sector,
leading to a massless down quark.
Thus,
this case is also ruled out.

\subsection{Flavour-changing Higgs couplings}

Here, we comment briefly on the appearance of
Flavour Changing Neutral Scalar Interactions (FCNSI).
As is well known,
large FCNSI with small scalar masses are precluded by measurements in the
neutral meson systems.
There are several ways to solve this problem.
One route is to endow the heavy neutral states with large masses.
This is achieved in the decoupling limit, with the added bonus
that it implies the alignment limit $\beta \sim \alpha + \pi/2$,
which is necessary to conform with the LHC measurements
of the $125$ GeV scalar.
Since this is always a possibility,
FCNSI cannot be used to exclude models from the
realm of possibilities.

A second route,
introduced by Glashow and Weinberg \cite{GW}
and independently by Paschos \cite{Paschos},
is to impose a $Z_2$ symmetry
under which all fields of equal charge transform equally,
thus implying vanishing FCNSI.
Neglecting right-handed neutrinos,
there are only four such possibilities -- for a review,
see, for example, Refs.~\cite{hhg, ourreview}.
When all fields are in $S_3$ singlets,
the situation reduces to $Z_2$.
When all fields of equal charge transform equally,
there are no FCNSI.
The cases where some fields of equal charge
transform differently were discussed in Ref.~\cite{Ferreira:2010ir}
and do lead to FCNSI.
Similarly,
when there are $S_3$ doublets in all sectors,
we do have FCNSI.

A third route is that the FCNSI are accidentally small.
This may be theoretically displeasing, but cannot be excluded.
A fourth route,
discussed by Branco, Lavoura, and Grimus (BGL),
is that the FCNSI are small because they are related
with the CKM matrix elements \cite{Branco:1996bq}.
We have checked that the models with $S_3$ doublets in all sectors
are not of the BGL type \cite{botella}.

\section{\label{sec:conclusions}Conclusions}

We study all representation assignments of $S_3$ in the 2HDM with quarks,
consistent with the basis requirements of non-vanishing, non-degenerate
masses, non-block diagonal CKM matrix and the presence of a CP violating phase.
We found that there are only two implementations consistent with this simple
requirements: all fields are in singlets or, else, all fields sectors
have a doublet representation.

When the scalars are in an $S_3$ doublet,
one must introduce soft-breaking terms.
Even with the most general real soft-breaking term,
there is a relation between $\alpha$ and $\beta$,
shown in Eq.~\eqref{new}.
As far as we know, this is a new result.

Finally, we point out that the viable models with doublet
representations are not of the BGL
type \cite{Branco:1996bq},
where the flavour changing scalar interactions are naturally
small because they are related to the CKM matrix elements
\cite{botella}.
It was shown in Ref.~\cite{Ferreira:2010ir}
that the only models based on abelian symmetries which
have this property are of the
type introduced originally in Ref.~\cite{Branco:1996bq}.
We now know that the non-abelian $S_3$ group does not
provide a further example of a BGL model.

\vspace{1ex}

\begin{acknowledgments}
We are grateful to I. Ivanov for discussions and to D. Emmanuel-Costa
for carefully reading the manuscript and for useful suggestions.
This work is supported in part by the Portuguese
\textit{Funda\c{c}\~{a}o para a Ci\^{e}ncia e Tecnologia} (FCT)
under contract UID/FIS/00777/2013.
D.~C. is partly supported by the Brazilian National Council
for Scientific and Technological Development (CNPq),
under grants 484157/2013-2 and 201066/2015-7
\end{acknowledgments}



\vspace{2ex}


\begin{thebibliography}{99}
%
\bibitem{ATLASHiggs}
G.~Aad {\it et al.}  [ATLAS Collaboration],
``Observation of a new particle in the search for
the Standard Model Higgs
boson with the ATLAS detector at the LHC,''
Phys.\ Lett.\ B {\bf 716}, 1 (2012)
[arXiv:1207.7214 [hep-ex]].
%
\bibitem{CMSHiggs}
S.~Chatrchyan {\it et al.}  [CMS Collaboration],
``Observation of a new boson at a mass of 125 GeV with the CMS
experiment at the LHC,''
Phys.\ Lett.\ B {\bf 716}, 30 (2012)
[arXiv:1207.7235 [hep-ex]].
%
\bibitem{hhg}
  J.F.~Gunion, H.E.~Haber, G.L.~Kane and S.~Dawson,
  \textit{The Higgs Hunter's Guide}
  \mbox{(Westview Press, Boulder, CO, 2000)}.
%
\bibitem{ourreview}
G.~C.~Branco, P.~M.~Ferreira, L.~Lavoura, M.~N.~Rebelo, M.~Sher, and J.~P.~Silva,
\emph{Theory and phenomenology of two-Higgs-doublet models},
\emph{Phys.\ Rept.\ }  {\bf 516}, 1 (2012)
[arXiv:1106.0034 [hep-ph]].
%
\bibitem{Ivanov:2007de}
I.~P.~Ivanov,
``Minkowski space structure of the Higgs potential in 2HDM.
II. Minima, symmetries, and topology,''
Phys.\ Rev.\ D {\bf 77}, 015017 (2008)
[arXiv:0710.3490 [hep-ph]].
%
\bibitem{Ferreira:2009wh}
P.~M.~Ferreira, H.~E.~Haber and J.~P.~Silva,
``Generalized CP symmetries and special regions of
parameter space in the two-Higgs-doublet model,''
Phys.\ Rev.\ D {\bf 79}, 116004 (2009)
[arXiv:0902.1537 [hep-ph]].
%
\bibitem{Ferreira:2010ir}
P.~M.~Ferreira and J.~P.~Silva,
``Abelian symmetries in the two-Higgs-doublet model with fermions,''
Phys.\ Rev.\ D {\bf 83}, 065026 (2011)
[arXiv:1012.2874 [hep-ph]].
%
\bibitem{Ferreira:2010bm}
P.~M.~Ferreira and J.~P.~Silva,
``A Two-Higgs Doublet Model With Remarkable CP Properties,''
Eur.\ Phys.\ J.\ C {\bf 69}, 45 (2010)
[arXiv:1001.0574 [hep-ph]].
%
\bibitem{Kajiyama:2013sza}
Y.~Kajiyama, H.~Okada and K.~Yagyu,
``Electron/Muon Specific Two Higgs Doublet Model,''
Nucl.\ Phys.\ B {\bf 887}, 358 (2014)
[arXiv:1309.6234 [hep-ph]].
%
\bibitem{Johansen:2015nxa}
A.~R.~Johansen and M.~Sher,
``Electron/muon specific two Higgs doublet model
at $e^{+}e^-$ colliders,''
Phys.\ Rev.\ D {\bf 91}, 054021 (2015)
[arXiv:1502.00516 [hep-ph]].
%
\bibitem{Ma:2004pt}
E.~Ma,
``Non-Abelian discrete family symmetries of leptons and quarks,''
hep-ph/0409075.
%
\bibitem{Ma:2013zca}
E.~Ma and B.~Melic,
``Updated $S_{3}$ model of quarks,''
Phys.\ Lett.\ B {\bf 725}, 402 (2013)
[arXiv:1303.6928 [hep-ph]].
%
\bibitem{Deshpande:1977rw}
N.G.~Deshpande and E.~Ma,
``Pattern Of Symmetry Breaking With Two Higgs Doublets,''
Phys.\ Rev.\  D {\bf 18}, 2574 (1978).
%
\bibitem{david}
D.\ Emmanuel-Costa,
private communication.
%
\bibitem{Nebot:2015wsa}
M.~Nebot and J.~P.~Silva,
``Self-cancellation of a scalar in neutral meson mixing and
implications for the LHC,''
Phys.\ Rev.\ D {\bf 92}, no. 8, 085010 (2015)
[arXiv:1507.07941 [hep-ph]].
%
\bibitem{Ja}
C.\ Jarlskog,
``Commutator of the Quark Mass Matrices in the Standard
Electroweak Model and a Measure of Maximal CP Violation,''
Phys.\ Rev.\ Lett.\ \textbf{55}, 1039 (1985);
``A Priori Definition of Maximal CP Violation,''
I.\ Dunietz, O.\ W.\ Greenberg, and D.-D.\ Wu,
Phys.\ Rev.\ Lett.\ \textbf{55}, 2935 (1985);
F.\ J.\ Botella and L.-L.\ Chau,
``Anticipating the Higher Generations of Quarks from
Rephasing Invariance of the Mixing Matrix,''
Phys.\ Lett.\ B {\bf 168}, 97 (1986);
J.\ Bernab\'eu, G.\ C.\ Branco, and M.\ Gronau,
``Cp Restrictions On Quark Mass Matrices,''
Phys.\ Lett.\ B \textbf{169}, 243 (1986).
%
\bibitem{GW}
S.~L.~Glashow and S.~Weinberg,
``Natural Conservation Laws for Neutral Currents,''
Phys.\ Rev.\ D {\bf 15}, 1958 (1977).
%
\bibitem{Paschos}
E.~A.~Paschos,
``Diagonal Neutral Currents,''
Phys.\ Rev.\ D {\bf 15}, 1966 (1977).
%
\bibitem{Branco:1996bq}
  G.~C.~Branco, W.~Grimus and L.~Lavoura,
  Phys.\ Lett.\ B {\bf 380}, 119 (1996)
  [hep-ph/9601383].
%
\bibitem{botella}
We are gratefull to F.~Botella for discussions
on this point.
%
%
%
%
%
\end{thebibliography}
\end{document}